\documentclass[twocolumn,secnumarabic,amssymb, nobibnotes, aps, prd]{revtex4-1}
\usepackage{graphicx}
\usepackage{dcolumn}
\usepackage{bm}
\usepackage{amsmath}
\usepackage{subfigure}%
\usepackage{color}

\begin{document}

\title{Effect of line defects on the electrical transport properties of monolayer MoS$_{2}$ sheet}%

\author{Amretashis Sengupta$^{1*}$, Dipankar Saha$^{1}$, Thomas A. Niehaus$^{2}$, Santanu Mahapatra$^{1}$}
\affiliation{$^{1}$Nano Scale Device Research Laboratory, Department of Electronic Systems Engineering,
Indian Institute of Science, Bangalore 560012, India\\ $^{2}$Institute I - Theoretical Physics, University of Regensburg, 93040 Regensburg, Germany}
\email{Corresponding Author: amretashis@gmail.com}
\date{\today}

\begin{abstract}
We present a computational study on the impact of line defects on the electronic properties of monolayer MoS$_{2}$. Four different kinds of line defects with Mo and S as the bridging atoms, consistent with recent theoretical and experimental observations are considered herein. We employ the density functional tight-binding (DFTB) method with a Slater-Koster type DFTB-CP2K basis set for evaluating the material properties of perfect and the various defective MoS$_{2}$ sheets. The transmission spectra is computed with a DFTB-Non-Equilibrium Green’s Function (NEGF) formalism. We also perform a detailed analysis of the carrier transmission pathways under a small bias and investigate the phase shifts in the transmission eigenstates of the defective MoS$_{2}$ sheets. Our simulations show a 2-4 folds decrease in carrier conductance of MoS$_{2}$ sheets in the presence of line defects as compared to that for the perfect sheet.
\end{abstract}
\maketitle
\section{Introduction}
Two dimensional (2-D) transition metal dichalcogenides, especially MoS$_{2}$ has shown great promise as an alternate channel material for next generation electron devices [1]-[2] due to their intrinsic non-zero band gap, which gives them a distinct advantage over graphene for logic circuit applications. Apart from this, 2-D materials offer enhanced electrostatic integrity, optical transparency and mechanical flexibility. Thus a great activity, in both experimental and computational research, has started to understand the electrical properties of atomically thin, layered MoS$_{2}$ crystals [1]-[8].\\
Although a number of studies have been conducted on possible enhancement/ degradation of MoS$_{2}$ electrical properties, not much focus has been given to defects and deformations in MoS$_{2}$ sheets. Point defects, dislocations, grain boundary effects can significantly impact the carrier transport in 2-D MoS$_{2}$ channels. Intrinsic crystallographic faults like line defect in single layer MoS$_{2}$ can occur due to stoichiometry changes in the S shell during fabrication. Recently Enyashisn et. al. [6] have reported experimental results on the various type of line defects present in 2-D  MoS$_{2}$. However their report is more focussed on defect formation and dynamics, rather than examining how such defects could alter the electrical properties of MoS$_{2}$ sheets.\\
This paper reports a computational study on the impact of such line defects on the electrical transport properties of monolayer MoS$_{2}$ sheets. We investigate four different types of line defects with the density functional tight-binding (DFTB) method in QuantumWise ATK.  We compute the density of states (DOS), transmission spectra, transmission eigenstates and the pathways for electron and hole transport in such defective and perfect sheets.\\
\section{Methodology}
Fig. 1 shows the various line defects in sheets of 2-D MoS$_{2}$ considered for our studies. We consider a monolayer MoS$_{2}$ sheet having length of 5 nm and width of 2.5 nm. Such supercell dimensions are considered sufficient for accurately simulating the electronic properties of a periodic sheet of MoS$_{2}$ as reported by other groups [9]-[12].
As shown in Fig. 1, the first type of defect {henceforth referred to as defect-I} has two zigzag edges joined in a bridging network of both S and Mo atoms. Defects-II and III have only Mo, as the bridging atom. Defect-III, differs from defect-II in the sense of the presence of an additional almost Stone-Wales type defect in the bridging Mo atoms alternating honeycombs. Defect IV consists of only S as the bridging atom and does not display any other variants as in defect-III. The defects are consistent with the results of Enyashin et. al.[6]\\
\begin{figure}[h]
\begin{center}
\includegraphics[width=0.9\columnwidth]{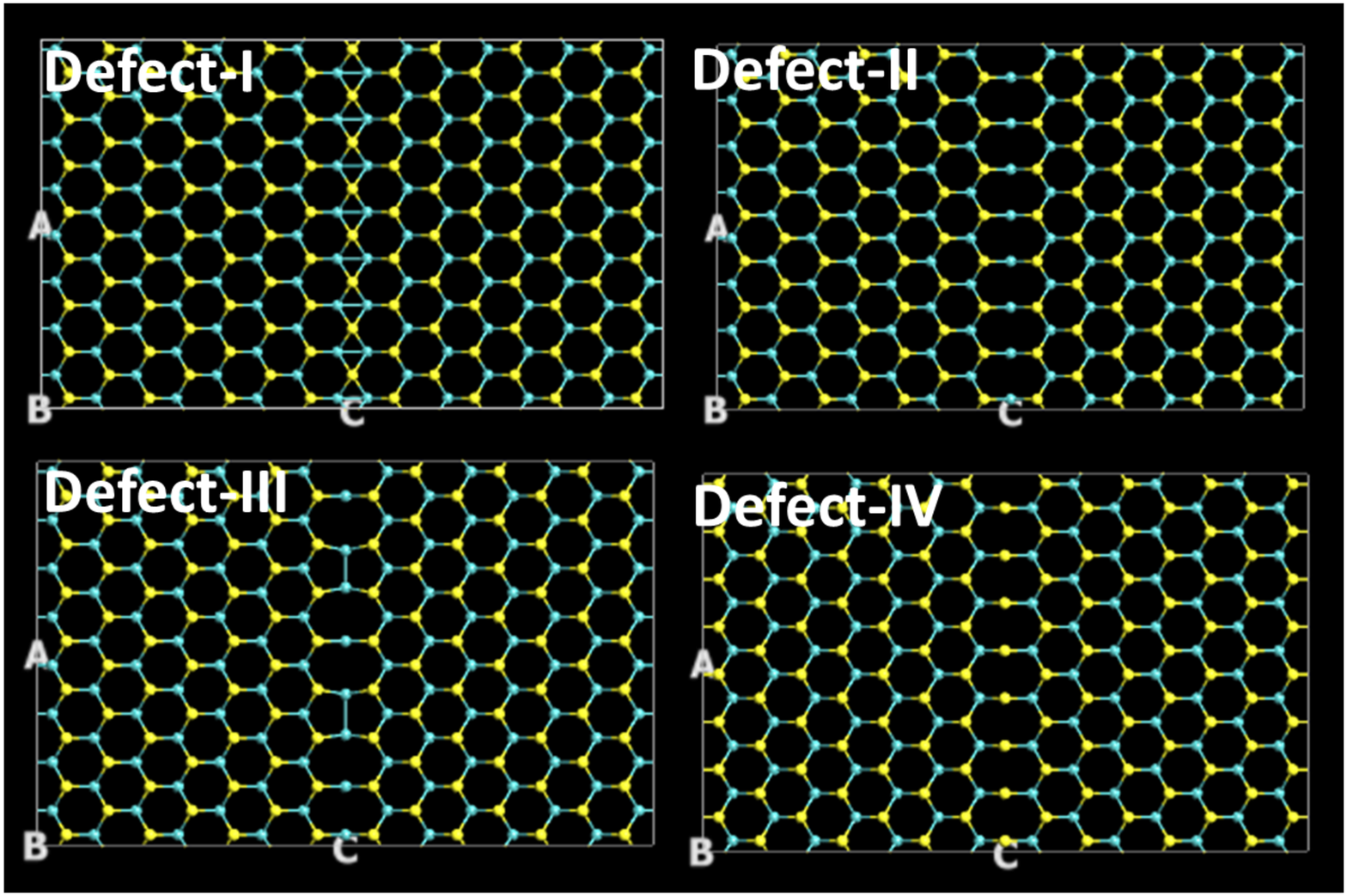}
\caption{The various line defects in 2-D MoS$_{2}$ considered in our study (top view).}
\label{fig_1}
\end{center}
\end{figure}
To calculate the electrical properties of our supercell we employ density functional tight binding (DFTB) theory in QuantumWise ATK 13.8.0  [13]. We use a 9$\times$1$\times$9 Monkhorst-Pack  k-grid[14] and employ the Slater-Koster type DFTB-CP2K basis set for MoS$_{2}$ available within the ATK package [15,16]. The transmission spectra and associated parameters such as the transmission eigenstates and the transmission pathways are simulated in Quantum Wise ATK using the self-consistent DFTB-NEGF method. This DFTB-NEGF method is chosen for our case as it has been found to yield fast and acceptably accurate simulations for large supercells of transition metal dichalcogenide materials [8]-[12].In the NEGF method we proceed to solve the Poisson- Schrödinger equation of the supercells self-consistently. It is assumed that the electrode regions are perfectly ohmic. Setting up the self-energy matrices $\Sigma_{L}$  and $\Sigma_{R}$ for the left and right contacts, the Green$'$s function $G$ is constructed as\\
\begin{equation}
G(E) = [EI-H-\Sigma_{S}-\Sigma_{L}-\Sigma_{R}]^{-1}
\label{equn_1}
\end{equation}
In (1) $I$ is the identity matrix and $\Sigma_{S}$ is the elastic carrier scattering self-energy matrix. From (1) parameters like the broadening matrices $\wp_{L}$ and $\wp_{R}$ and the spectral densities $A_{L}$ and $A_{R}$ are evaluated using the following relations\\
\begin{equation}
\wp_{L,R}=i[\Sigma_{L,R}-\Sigma^{\dagger}_{L,R}]
\label{equn_2}
\end{equation}
\begin{equation}
A_{L,R}=G(E)\wp_{L,R}G^{\dagger}(E)
\label{equn_3}
\end{equation}
The density matrix [$\Re$] used to solve the Poisson equation is given by
\begin{equation}
[\Re] = \int_{-\infty}^\infty{\frac{dE}{2\pi}[A(E_{k,x})]f_0(E_{k,x}-\eta)}
\label{equn_4}
\end{equation}
where $A(E_{k,x})$ is the spectral density,$E_{k,x}$ the energy of the conducting level,and $\eta$ being the chemical potential of the contacts. $f_{0}(.)$ is the Fermi function. For NEGF studies, a multi-grid poisson solver is employed using Dirichlet boundary conditions on the left and right faces (i.e. the electrodes) and periodic boundary conditions along the width of the supercell. The electrode temperatures are considered 300 K. The carrier density $n_{tot}$ is  evaluated from the NEGF formalism and put into the Poisson solver to self-consistently and iteratively evaluate the potential $U_{SCF}$ . The converged values of the carrier density and the self-consistent potential are used to calculate the transmission matrix $\Im(E,V)$ 
\begin{equation}
\Im(E,V)=trace[A_{L}\wp_{R}]=trace[A_{R}\wp_{L}]
\label{equn_5}
\end{equation}
For the transmission spectra we use the Krylov self-energy calculator [13] with the average Fermi level being set as the energy zero parameter.\\
The transmission pathways are evaluated by splitting the transmission coefficient into local bond contributions $\Im_{i,j}$. The pathways are such that, if the system is divided into 2 parts $a$ and $b$ then the pathways across the boundary between $a$ and $b$ sum up to the total transmission coefficient as [13,17,18]\\
\begin{equation}
\Im  = \sum\limits_{i \in a,i \in b} {{\Im _{i,j}}}
\end{equation}
The local bond contributions, $\Im_{i,j}$ , can be both positive and negative. A negative value correspond to that the electron is back scattered along the bond. We plot the relative local transmission with coloured arrows with a threshold value of 0.05 $\AA$ . The transmission is scaled as $\frac{t}{t_{max}}\times p$, $p$ being the scaling power. The value of $p$ is taken to be 0.01.\\ 
To find the transmission eigenstates of the sheet, the transmission matrix can be written as 
\begin{equation}
\Im _{nm} = \sum\limits_k {{t_{nk}}t_{km}^\dag}
\end{equation}
Where $t_{nk}$ is the transmission amplitude from Bloch state   $\psi_{n}$ in the left electrode to Bloch state $\psi_{k}$ in the right electrode. The transmission coefficient is given by the trace of the transmission matrix $\Im  = \sum\limits_n {{\Im _{nn}}} $. The transmission eigenstates are obtained by propagating the linear combination of the Bloch states
\begin{equation}
\sum\nolimits_n {{\ell _{\alpha ,n}}{\psi _n}}
\end{equation}
where ${\ell _{\alpha ,n}}$  diagonalize the transmission matrix as
\begin{equation}
\sum\limits_m {{\Im _{nm}}} {\ell _{\alpha ,m}} = {\lambda _\alpha }{\ell _{\alpha ,n}}
\end{equation}
with transmission eigenvalue $\lambda_{\alpha}$.\\
\section{Results and discussions}
In Fig. 2 (a)-(b), we show the density of states (DOS) and the normalized transmission spectra of the perfect MoS$_{2}$ sheet under consideration. From Fig. 2 (a) a forbidden gap of 1.84 eV is estimated between the valence band (VB) maxima and conduction band (CB) minima of the material. This is in accordance with reported band gap of 1.78 eV of perfect MoS$_{2}$ sheets. [7]-[12] The most significant contributions to the DOS come from the p and d orbitals of Mo and S atoms. If we see the normalized transmission spectra for the perfect sheet as in Fig. 2(b), we find two distinct peaks at 1.5 eV and -1.68 eV. Upon the application of a small bias of 0.2 V (on either the left or right electrode keeping the other terminal ground) we observe the edges of the transmission spectra drawing slightly closer to the Fermi level. This signifies availability of a greater number of conducting channels both in CB and VB.\\
\begin{figure}[!htbp]
\subfigure[]{\includegraphics[width=0.45\columnwidth]{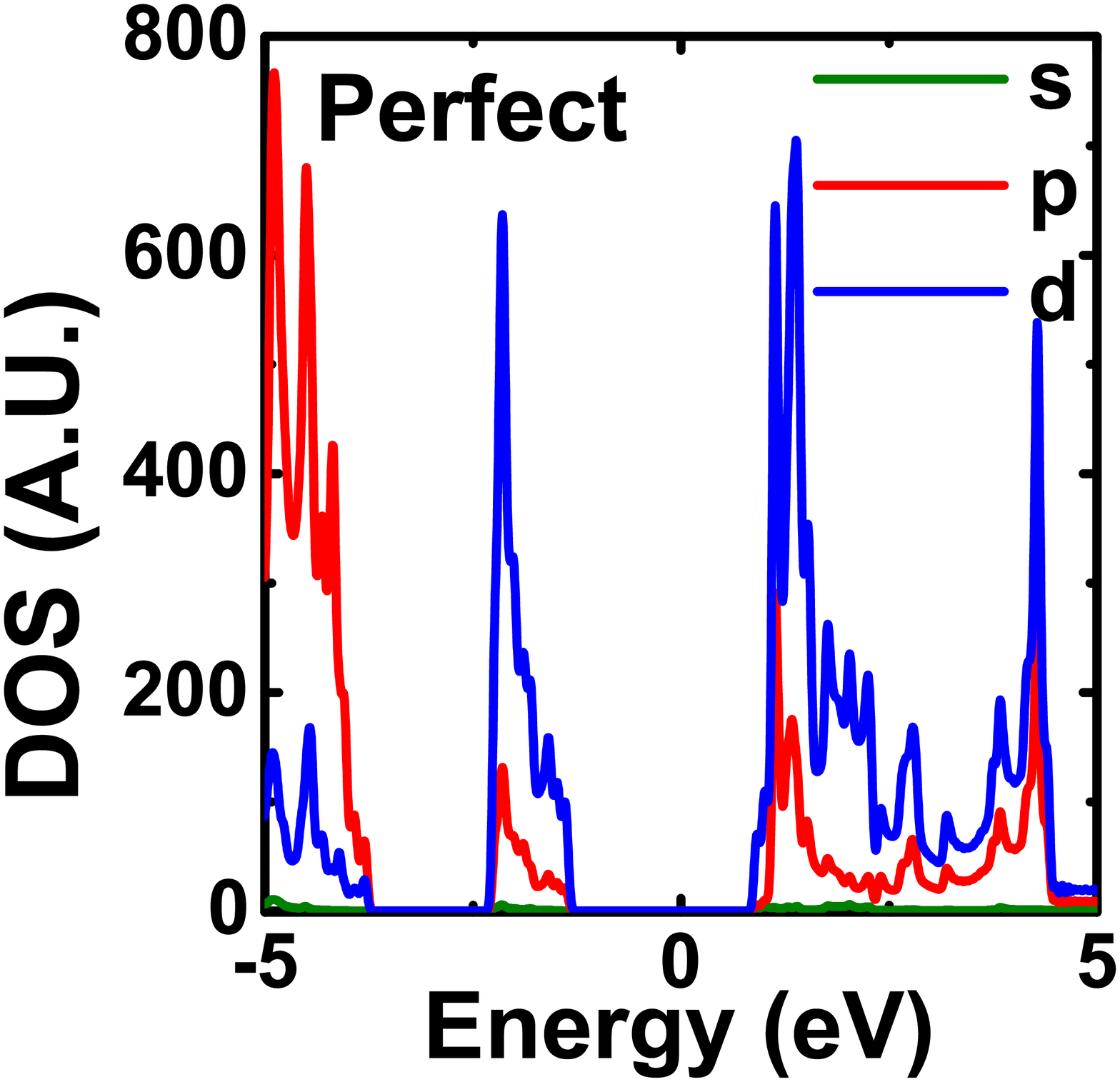}\label{a}}
\subfigure[]{\includegraphics[width=0.45\columnwidth]{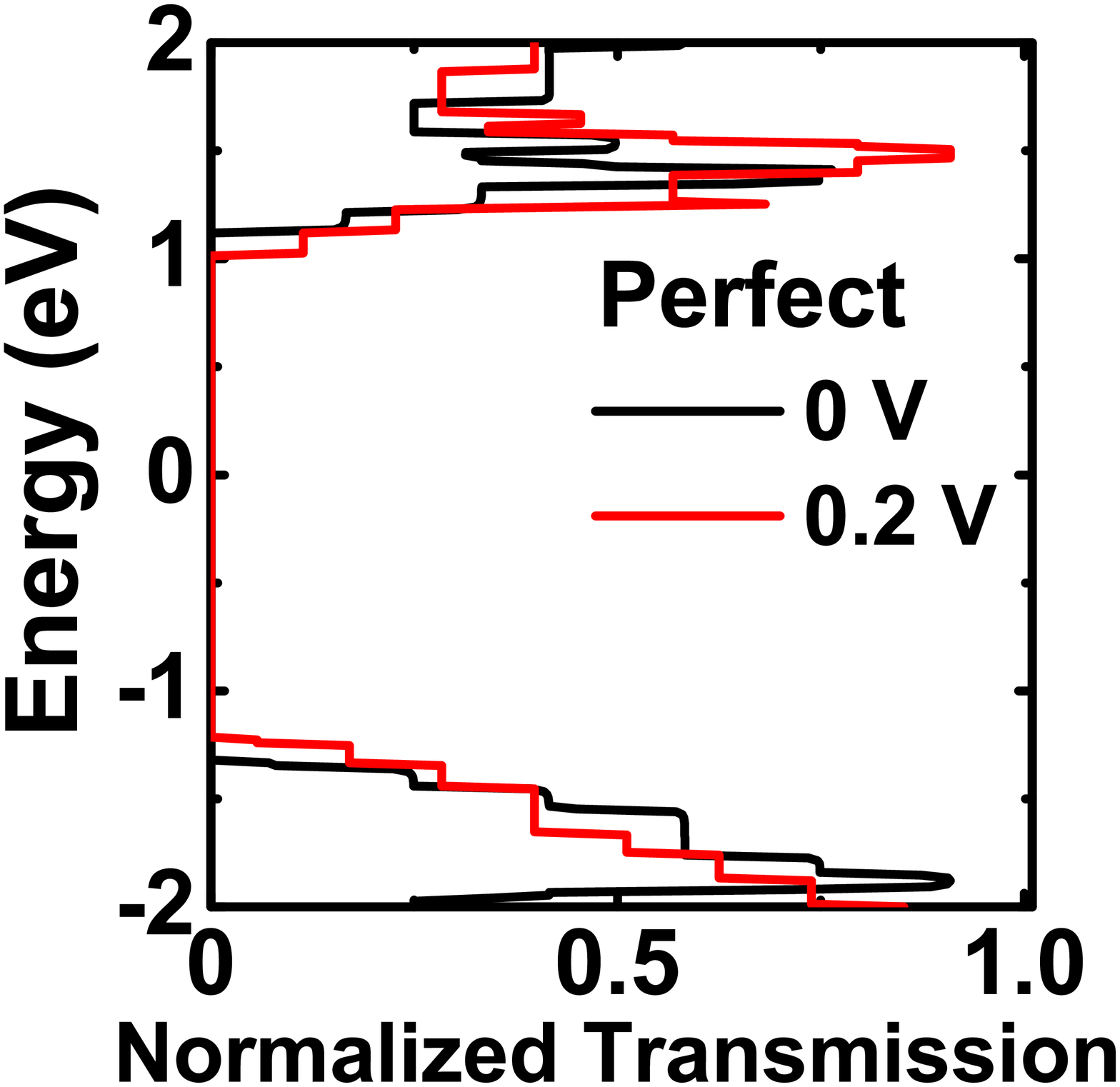}\label{b}}
\caption{(a) The density of states (DOS) and (b) Transmission spectra of perfect MoS$_{2}$ supercell}
\label{fig_2}
\end{figure}
In Fig 3, we have the DOS of the various defective sheets under consideration. Here we observe the states to be more evenly spread out, having much less sharper peaks in the DOS compared to perfect MoS$_{2}$ sheet. The defect states are seen to exist in considerable density even in the formerly forbidden gap of the perfect MoS$_{2}$. This indicates the possibility of a semiconductor to semi-metal transition (depending on the contribution of these defect states to the carrier transmittance) of the MoS$_{2}$ sheets in the presence of the line defects.\\
\begin{figure}[!htbp]
\begin{center}
\includegraphics[width=1.0\columnwidth]{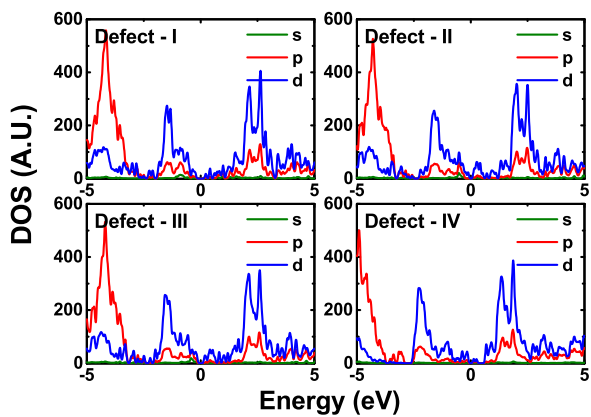}
\caption{Density of states (DOS) of various types of line defects in MoS$_{2}$ sheets. Defect states contributions are seen in the region near the Fermi level (set at zero).}
\label{fig_1}
\end{center}
\end{figure}
From the transmission spectra in Fig. 4 of the various defective sheets, we observe that the staircase like behavior of the spectra (as seen for the perfect sheet) no longer persists and the defect states near the Fermi level (as seen in the DOS plots) do not contribute much to the number of transmission channels. To study the reason for this behavior we conduct studies on the transmission eigenstates and the carrier transmission pathways of the different sheets. It should be noted that the transmission spectra shown in Fig. 4, has been normalized in order to gain a better understanding of the relative contribution of various energy bands to the total transmission spectra of the MoS$_{2}$ sheets. We show mostly the region near the Fermi level of the spectra, since for small bias this is the most significant part of the transmission spectra that contributes to carrier transport. With a small applied bias on either left/ right terminal of the sheets, the contributions of the various available channels seem to undergo some minor redistribution. Under such bias, the transmission peaks, signifying most transparent channels for carrier conduction for defect-I are observed at 1.33 and -1.88 eV. For defect-II the peaks are at 1.56 eV and -1.92 eV, for defect –III these are 1.54 eV, -1.72 eV and for defect –IV the peaks are at 1.54 eV and -1.74 eV respectively.\\
In our studies we consider the elastic scattering in the channel for electron/ hole transport. Therefore it is of interest to look into the transmission eigenstates and the transmission pathways for the defective sheets. In order to visualize the transmission eigenstates and pathways we select the energy values corresponding to maximum transmission (as in Fig. 4) for each case. For the electron transmission eigenstates and the transmission pathways the energy state corresponding to the transmission peak on the CB are considered.  For the hole transmission it is the energy state corresponding to the transmission peak on the VB that are considered.\\
\begin{figure}[!htbp]
\begin{center}
\includegraphics[width=1.0\columnwidth]{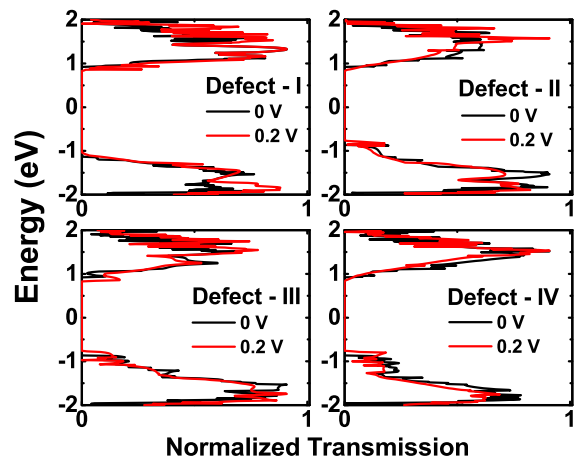}
\caption{Transmission spectra under zero-bias and a low applied bias (0.2 V) condition for the various types of line defects in MoS$_{2}$ sheets. The change in the transmission spectra is independent of the choice of bias terminal.}
\label{fig_1}
\end{center}
\end{figure}
\begin{figure*}[!htbp]
\subfigure[]{\includegraphics[width=1.0\columnwidth]{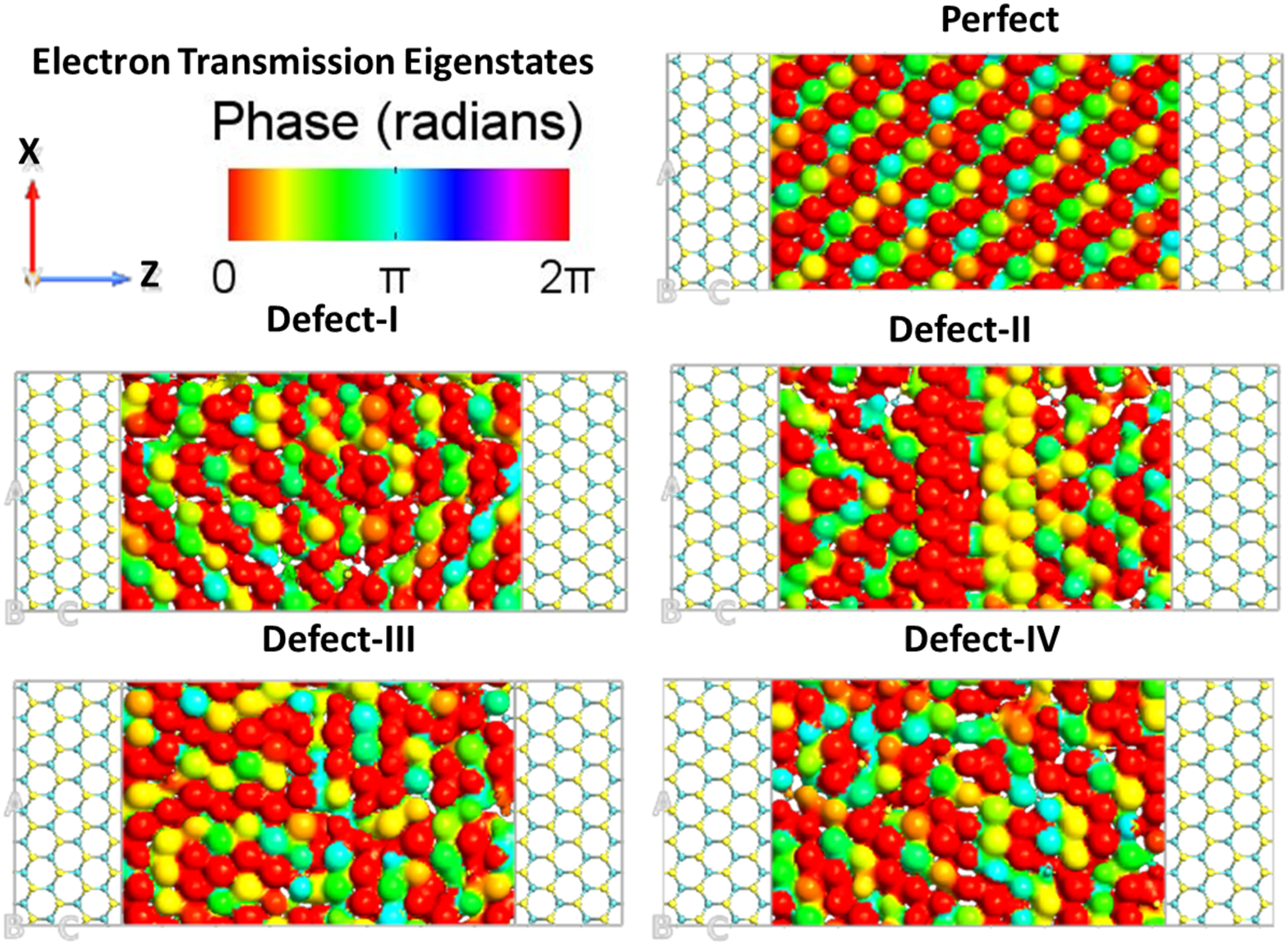}\label{a}}
\subfigure[]{\includegraphics[width=1.0\columnwidth]{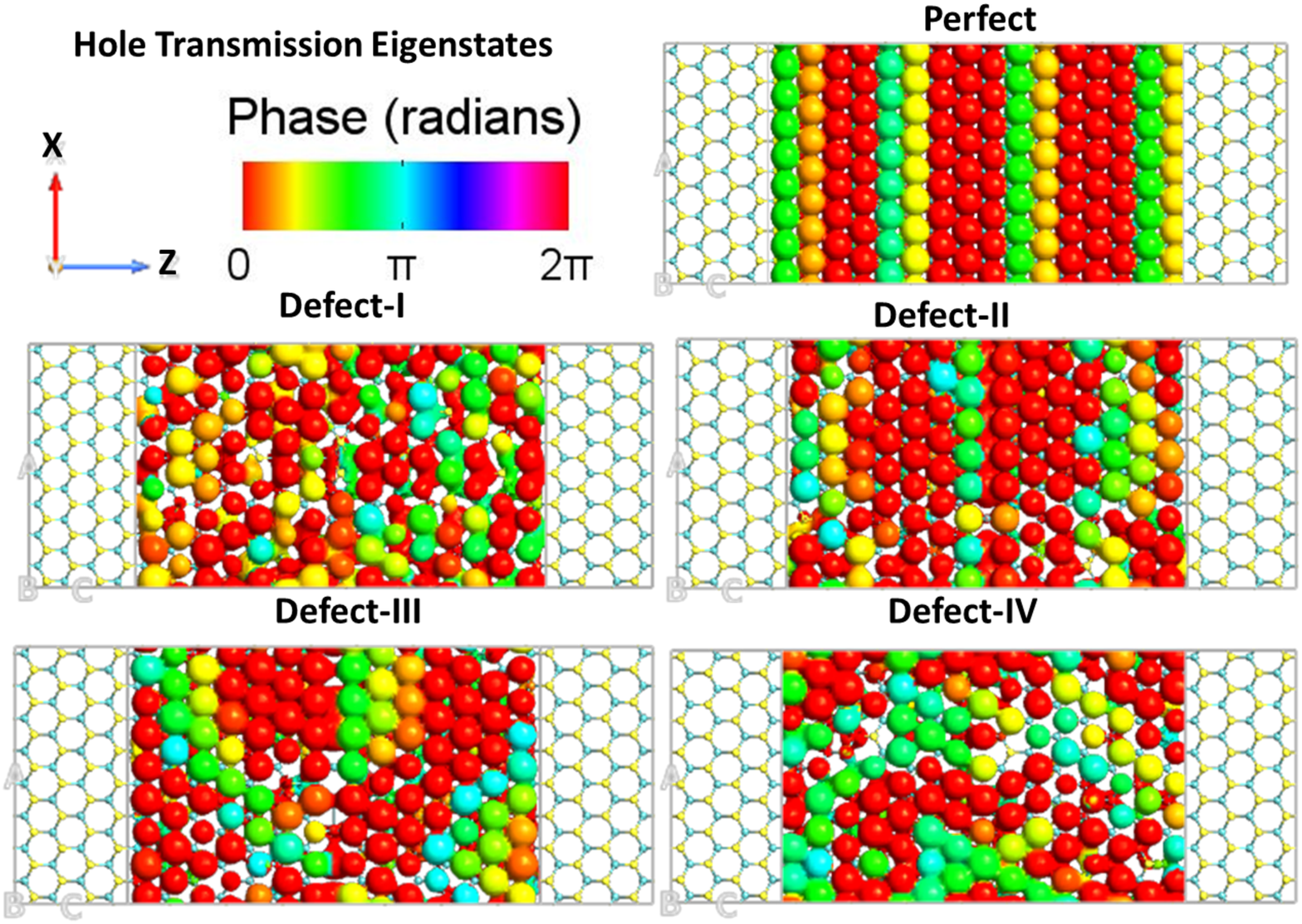}\label{b}}
\caption{(a) Electron and (b) Hole transmission eigenstates isosurface plots for the perfect and the various defective sheets. Isosurfaces are plotted for an applied bias of 0.2V, at the CB transmission peak for electrons and VB transmission peak for holes.}
\label{fig_2}
\end{figure*}
\begin{figure*}[!htbp]
\subfigure[]{\includegraphics[width=1.0\columnwidth]{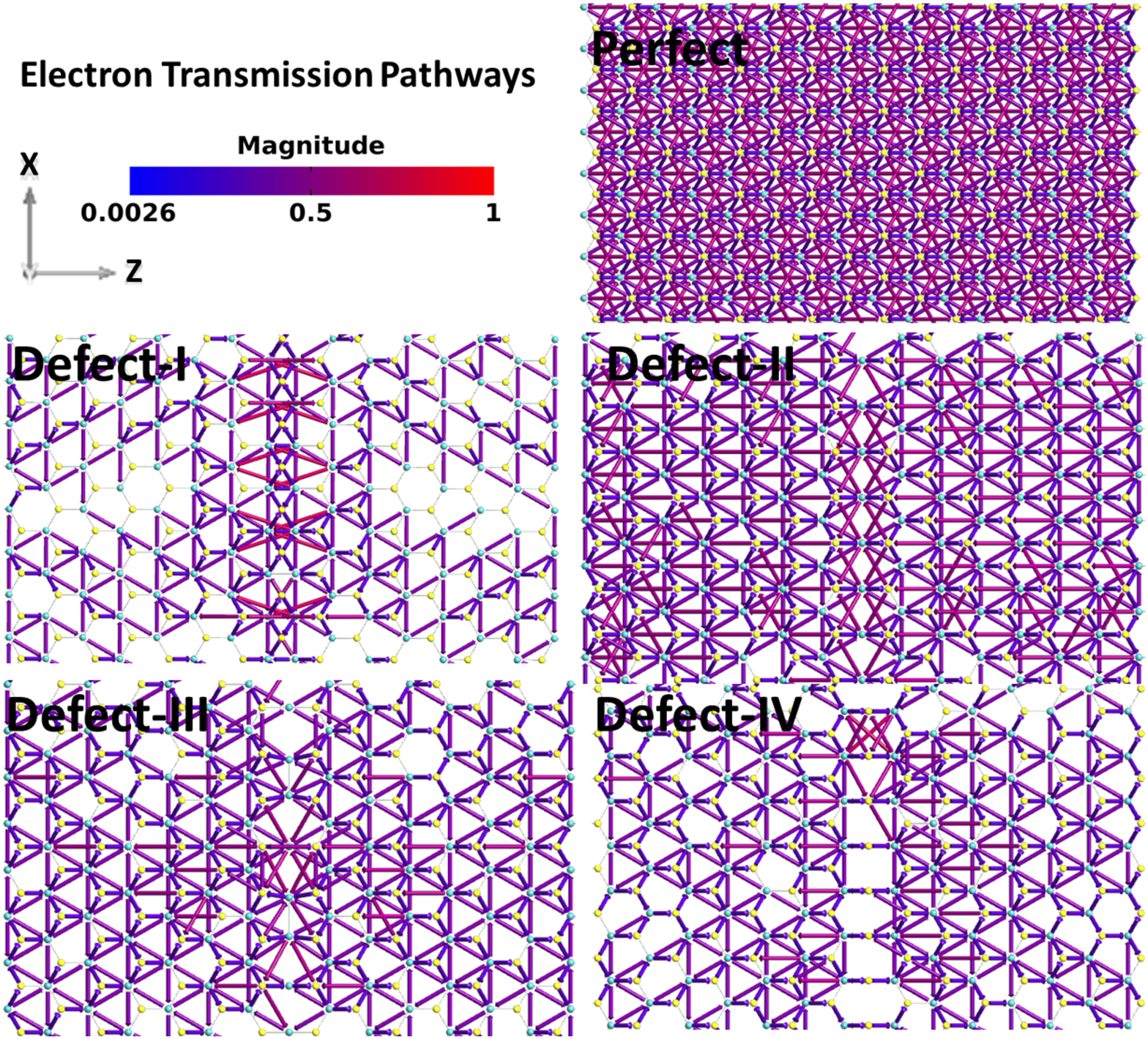}\label{a}}
\subfigure[]{\includegraphics[width=1.0\columnwidth]{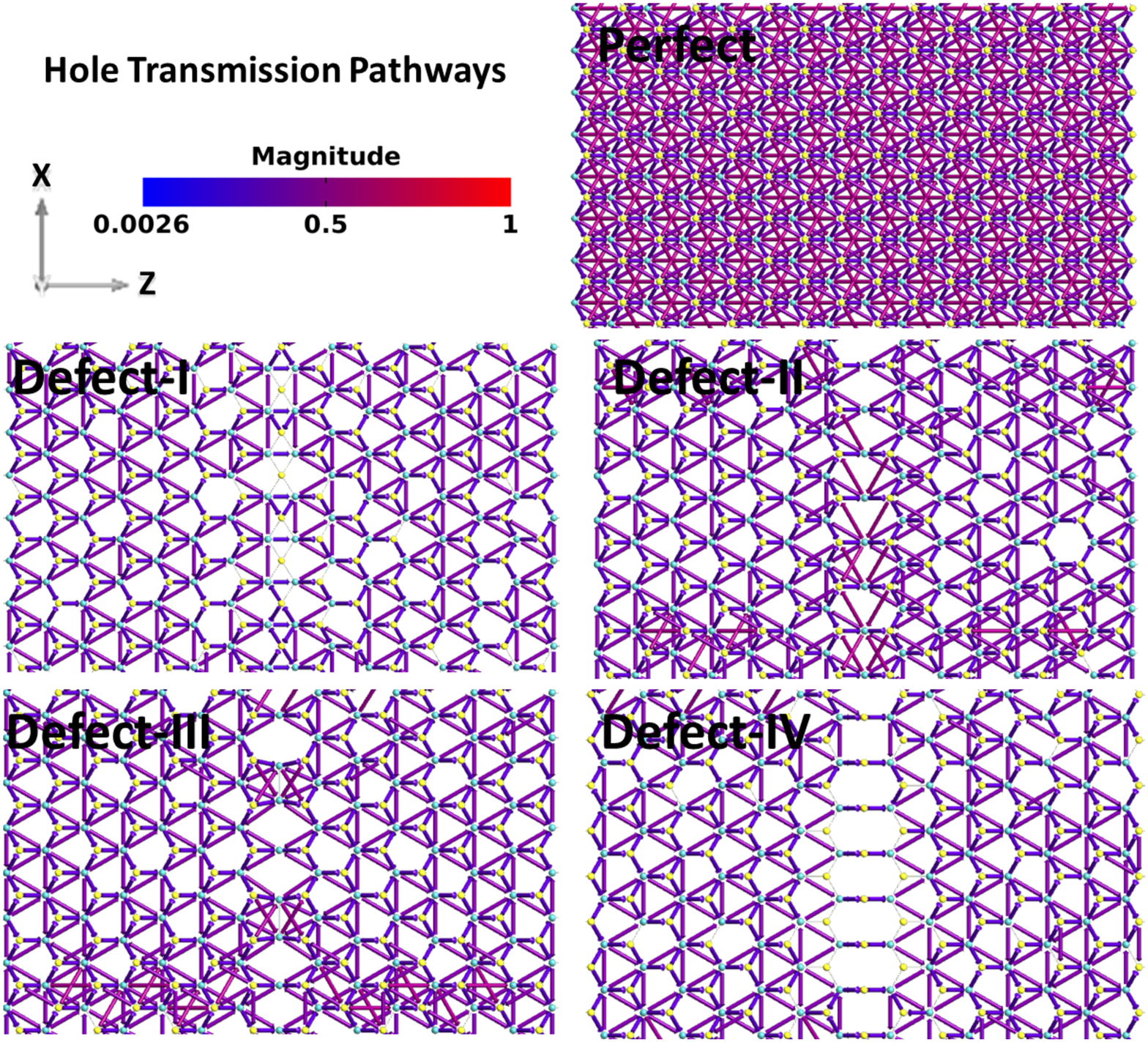}\label{b}}
\caption{(a) Electron and (b) Hole transmission pathways plotted for an applied bias of 0.2 V, at the CB transmission peak for electrons and VB transmission peak for holes.}
\label{fig_2}
\end{figure*}
From Fig. 5 (a)-(b), we see that there exists a certain periodic pattern in the phase of the transmission eigenstates for the perfect case. In the case of the line defects this periodic pattern is no longer preserved and a varying degree of randomness in phase sets in for both the electron and the hole transmission eigenstates.  Among the defects in case of defect-II, there exists an interesting pattern of about $\frac{\pi}{2}$  phase shift of the peak transmission eigenstate across the line defect for electron transport. Also for the hole transmission eigenstate, there exist about $\frac{3\pi}{4}$ phase difference across the line defect in defect –II. It is defect-I that seems to have most random variance of phase, followed by defect-IV and III. The high degree of randomization in the phase of the transmission eigenstates due to the defects, can signify lesser transmission across the defective sheets, and suppress conductance of the defective sheets.\\
Fig. 6 (a)-(b), shows the electron and hole transmission pathways for the various systems under consideration. The relative scale factor and the threshold values are as mentioned in the methodology section, and they are kept the same for all the pathway plots. From the plots we see that the pathways for electron and hole transmission at the respective CB and VB transmission peaks are identical in case of perfect sheet of MoS$_{2}$. This is in accordance with the closely matching values of electron and hole effective masses for perfect MoS$_{2}$ sheets as reported in our earlier work.[19] However for the line defects, the transmission pathways differ significantly for electrons and holes.\\
In defect-I, the number of available transmission paths near the defect site seems more for electron transport compared to that for hole transport, also the magnitude of transmission through these pathways is higher compared to that for the holes. In defect-I the transmission path between the Mo atoms around the line defect has the highest magnitude of electron conductance as seen in Fig. 6(a). The same pathways do not seem equally conducive to hole transport, as evident from Fig. 6(b). However away from the defect site, there exist lesser number of paths for electrons than that for holes in defect-I.\\
For defect-II also, there exist a larger number of paths near the defect for electron transport than that for hole transport. Same is observed for defects-III and IV. Overall in all the defects we observe some highly electron conducting paths between the metallic atoms (Mo-Mo) around the defects. However for hole transport this route does not offer any enhanced carrier transport and is equivalent in transport magnitude to the Mo-S pathway in case of holes. The S-S pathway is weakly conducive in either electron or hole transport.\\
\begin{figure}[!htbp]
\includegraphics[width=0.75\columnwidth]{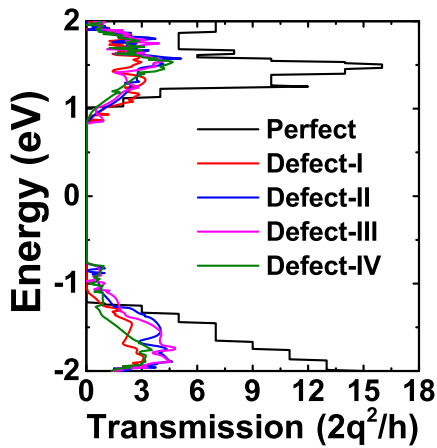}
\caption{Absolute transmission under low-bias (0.2 V) conditions}
\label{fig_2}
\end{figure}
In Fig 7, we show the absolute value of transmission spectra of the perfect and the various defective sheets under a small applied bias of 0.2 V. It is worth mentioning here that the transmission spectra shown earlier in Fig. 2(b) and Fig. 4, are normalized to provide a better view of how the relative contribution of various energy bands to the total transmission of the channel (MoS$_{2}$ sheet) evolve under applied bias. In order to understand the carrier conductance through the channel, it is the absolute value of the transmission that is more significant than the normalized $\Im(E,V)$.
As seen in the Fig 7 the magnitude of transmission through the perfect sheet is significantly larger than that for the defective ones. The defect states have a slight contribution to transmission near the energy levels of 1 eV and -1 eV, whereas the transmission for the perfect sheet in these regions is somewhat lesser. Also for the defective sheets the levels are less discrete and do not have the staircase behavior as in the perfect sheet. At the corresponding transmission peaks, the value of $\Im(E,V)$ for the perfect sheet is 2$-$4 times that of the various defective sheets. This indicates a high degree of suppression of carrier transmission in presence of the line defects in MoS$_{2}$ sheet.\\
\section{Summary}
Here we report the impact of four different types of line defects on the electrical transport properties of monolayer MoS$_{2}$ sheets. The defects with different configurations and Mo and S as the bridging atoms, based on recent experimental evidence, are studied with the DFTB-NEGF formalism for their transport behavior.\\
Our studies show that although a moderate amount of defect states are induced near the Fermi level due to these line-defects, the carrier transport through these defect states is rather limited and the overall transmission is highly scattered in presence of these line defects. The suppressed carrier transmission is also reflected in the high degree of defect induced randomization in the phase of the transmission eigenstates. The lowering of transmission in the defective sheets and the elastic scattering at the defect centres were evident from the transmission pathways. Our studies show a sizeable decrease in carrier conductance in MoS$_{2}$ sheets due to the presence of line defects as compared to that for the perfect sheet. Such results are very significant considering the future application of MoS$_{2}$ monolayer sheets for MOSFET applications.

\begin{acknowledgments}
Dr. A. Sengupta thanks DST, Govt. of India, for the DST Post-doctoral Fellowship in Nano Science and Technology.This work was supported by the Department of Science and Technology, Government of India, under grant no: SR/S3/EECE/0151/2012.
\end{acknowledgments}

\end{document}